\documentclass[12pt,preprint]{aastex}

\newcommand{\etal}{{et\thinspace al.} }
\newcommand{\Lya}{{$Ly\alpha$} }

\begin{document}

\title{The Major Sources of the Cosmic Reionizing Background at $z\simeq$ 6}

\author{Haojing Yan}
\affil{SIRTF Science Center, California Institute of Technology, MS 100-22,
Pasadena, CA 91125}
\email{yhj@ipac.caltech.edu}
\author{Rogier A. Windhorst}
\affil{Department of Physics \& Astronomy, Arizona State University, Tempe, AZ
85287}
\email{Rogier.Windhorst@asu.edu}

\begin{abstract}

In this paper, we address which sources contributed most of the reionizing 
photons. Our argument assumes that the reionization ended around $z\simeq 6$
and that it was a relatively quick process, i.e., that there was a 
non-negligible fraction of neutral hydrogen in the Universe at somewhat earlier
epochs. Starting from our earlier estimate of the luminosity function (LF) of
galaxies at $z\simeq 6$, we quantitatively show that the major sources of 
reionization are most likely galaxies with $L<L_*$. Our 
approach allows us to put stronger constraints to the LF of galaxies at 
$z\simeq 6$. To have the Universe completely ionized at this redshift, the 
faint-end slope of the LF should be steeper than $\alpha=-1.6$, which is the 
value measured at lower redshifts ($z\simeq 3$), unless either the
normalization ($\Phi_*$) of the LF or the clumping factor of the ionized 
hydrogen has been significantly underestimated. If $\Phi_*$ is actually lower
than what we assumed by a factor of two, a steep slope close to $\alpha=-2.0$
is required. Our LF
predicts a total of 50 -- 80 $z\simeq 6$ galaxies in the HST Ultra Deep Field
(UDF) to a depth of $AB=28.4$ mag, which can be used to constraint both
$\Phi_*$ and $\alpha$.
We conclude that the $least$ luminous galaxies existing at this redshift 
should reach as low as some critical luminosity in order to accumulate the
entire reionizing photon budget. On the other hand, the existence of
significant amounts of neutral hydrogen at slightly earlier epochs, e.g.
$z\simeq 7$, requires that the $least$ luminous galaxies should not be fainter
than another critical value (i.e., the LF should cut-off at this point). 
%The upcoming observation of the Ultra Deep Field (UDF) by the Hubble Space
%Telescope (HST) will be able to narrow down the possible range of the 
%faint-end slope of the LF, and hence constrain these thresholds. Such
%quantitative results will thus help us understand the star-forming objects
%in early Universe that caused reionization.

\end{abstract}


\section{Introduction}

    High resolution UV spectra of $z>5$ quasars found in the Sloan 
Digital Sky Survey (SDSS) provide solid evidence that the intergalactic
hydrogen was in a {\it completely ionized} state up to an early epoch of 
$z\simeq 6$ (e.g., Fan \etal 2001). At slightly higher redshifts, however,
a small fraction of neutral hydrogen is being seen, as indicated by the 
complete Gunn-Peterson trough found in several SDSS quasars at $z=6.28$ -- 
6.43 (Becker \etal 2001; Fan \etal 2003). This has been used to argue that $z\simeq 6$ marks the
end of the reionization era, and that the redshift at which the reionization 
began may not be much higher than $z\simeq 6$ (Becker \etal 2001; Fan \etal 2002).
Recently, Cen (2003) proposed that the Universe might have been
reionized twice, where the first epoch was completed at $z\simeq 15$--16 and the
second epoch at $z\simeq 6$. The first reionization is consistent with an
early beginning of reionization suggested by the recent Wilkinson Microwave
Anisotropy Probe (WMAP) results (e.g., Spergel \etal 2003),
while the second reionization is consistent with the SDSS results.

   It is still not certain what kind of objects provided the reionizing
background. Here we focus on the end of the reionization era at $z\simeq 6$. 
There seems to be a consensus that the AGN population is not sufficient to
account for the entire required reionizing photon budget (e.g., Fan \etal 2002),
which leaves $normal$ star-forming galaxies as the only alternative. The 
discovery of \Lya emitters at $z\simeq 6.5$ (Hu \etal 2002; Kodaira \etal 2003)
suggests that star-forming galaxies did exist at such early epochs, and the
significant number of faint Lyman-break galaxy (LBG) candidates at 
$z\simeq 6$--6.5 found by Yan \etal (2003a) suggest that these objects could 
indeed be responsible for the completion of the reionization at this epoch. In 
this letter, we further study the contribution of galaxies and AGN to the 
reionizing background, and point out 
that dwarf galaxies may well have contributed the vast majority of the 
reionizing photons. Our investigation provides significant constraints to the
luminosity function (LF) of galaxies at $z\simeq 6$, most importantly the 
faint-end slope of the LF, and the critical luminosity of the $least$ luminous 
galaxies that must have existed at $z\simeq 6$.

   This paper is organized as follows. In \S 2, we summarize our current best
knowledge of the LF of galaxies and AGN at $z\simeq 6$, which forms the base of
our discussion. The contribution of
these two types of objects to the total ionizing background is calculated in
\S 3, followed by a discussion on the implications of our result in \S 4. A
summary is given in \S 5. Throughout the paper, we adopt a cosmology of
$\Omega_M=0.27$, $\Omega_\Lambda=0.73$, and
$H_0=71$ km$\,$s$^{-1}\,$Mpc$^{-1}$.

\section{The Luminosity Functions of Galaxies and AGN at $z\simeq 6$}

   Currently, the number of confirmed galaxies at $z\simeq 6$ is still 
insufficient to derive their LF in a conventional way. Yan \etal (2002)
presented an $estimate$ of the LF of galaxies at $z\simeq 6$ in terms of their
cumulative surface density as a function of apparent magnitude. This estimate
was made by using a Schechter function,
\begin{displaymath}
\Phi(M)=0.921\cdot \Phi_*\cdot 10^{0.4(\alpha+1)(M_*-M)}\cdot exp\left[-10^{0.4(M_*-M)}\right],
\end{displaymath}
with the values of its three free parameters adopted as following: the 
characteristic absolute magnitude $M_*$ (measured around rest-frame 1300\AA)
and the faint-end slope $\alpha$ obtained at $z\simeq 3$ (Steidel \etal 1996) 
were assumed to be the same at $z\simeq 6$, and the scale factor $\Phi_*$ was 
found by using the cumulative number density of $z\geq 5.5$ galaxies detected
in the Hubble Deep Field North (HDF-N) as the normalization. In the 
cosmological model used in this paper, these three parameters are 
$M_{AB}=-21.03$ mag, $\Phi_*=4.55\times 10^{-4}$ Mpc$^{-3}$ mag$^{-1}$, and 
$\alpha=-1.6$. 
%In the search of $z\simeq 6$ objects with the deep data obtained
%by the Advanced Camera for Surveys (ACS) on-board the HST, Yan \etal (2003a)
%pointed out that the faint-end slope could be as steep as $\alpha \simeq 2.0$.
The predicted cumulative surface density based on this estimated LF is 
reproduced here in Fig. 1 (red curves) with a few revisions. 

   All observational constraints available to date are also plotted in this
figure for comparison. Among them, those based on the SDSS quasar hosts, the
Keck \Lya emitters and the deep ACS parallel field were discussed in
Yan \etal (2002; 2003a\footnote{Note that the survey depth quoted in Yan \etal
(2003a) is incorrect. These ACS parallel data were taken at a gain of
4 $e^-$/ADU, and we corrected for this before applying the magnitude
zeropoints that were measured at a gain of 1 $e^-$/ADU. However, this
correction turns out to be unnecessary, since the HST on-the-fly pipeline
already makes the extra step of dividing the gain value into the flat-fields,
which is not documented in the obvious places. As a consequence, the 
magnitudes of the $z\simeq 6$ candidates reported in Yan \etal (2003a) should
be 1.5 mag brighter than their quoted values. Fig. 1 has reflected this change,
which makes the ACS parallel upper limit less restrictive. Other than these, the
major conclusion in that paper remains unchanged. This change does not affect
this current paper.}).
The three new upper limits at the bright-end are
discussed in Yan (2003b). In short, the NOAO-4m MOSAIC LBG and \Lya limits are
derived from a degree-sized intermediate-band survey centered on the HDF-North
and South (Yan \etal 2003, in preparation), and the Subaru LBG limit is what
we derived based on the deep Subaru data of the HDF-N (Capak \etal 2003). The
data point marked as a cross is derived from the Large Area Lyman-Alpha (LALA)
survey 
(Rhoads \& Malhotra 2001; Rhoads \etal 2003). Recently, three more measurements
have become available. Stanway \etal (2003a, b) used the public Great 
Observatories Origins Deep Survey (GOODS) single-epoch data to search for 
$z\simeq 6$ objects. Bouwens \etal (2003) searched for such objects in their
deep ACS Guaranteed Time Obsevation (GTO) data. Most recently, the GOODS team
 released its
$z\simeq 6$ candidate list derived from the stacked three-epoch ACS data
(Dickinson \etal 2003). All these new observations agree with our $z\simeq 6$ LF.
%(the result of Stanway \etal will not be discussed separately, as it is 
%incorporated in the GOODS result (see the discussion of Dickinson \etal 2003).
%The data point of Bouwens \etal (2003) is shown as the solid triangle.
The GOODS candidate catalog has sufficient statistics to allow us to derive not
just one, but several constraints at different flux levels from 25.0 to 27.0
mag.
%Taking these candidates at face value, the inferred GOODS cumulative surface densities from 25.0 to 27.0 mag are shown as open pentagons in the figure. 
All these values agree very well with our LF from 23.0 to 26.5 mag. At 
$m_{AB} >26.5$ mag, the GOODS counts are significantly lower than our estimate,
which is possibly due to the incompleteness 
of the catalog at the faint-end. To summarize, our LF broadly agrees with all
known constraints, and is about the best that one can get with the available
data so far. 

    For our purpose, an analytic form of the AGN LF is preferred. The 
``standard" double power-law quasar LF (e.g., Boyle, Fong \& Shanks 1988)
is therefore used:
\begin{displaymath}
   \Phi(M) = \frac{\Phi_*}{10^{0.4(\beta_1+1)(M-M_*)}+10^{0.4(\beta_2+1)(M-M_*)}
}.
\end{displaymath}
This form has been proved to be a very good representation for quasars up to at
least $z=2.5$. There are four free parameters involved: the turn-over
magnitude $M_*$, the scaling factor $\Phi_*$, the bright-end slope $\beta_1$, 
and the faint-end slope $\beta_2$. The SDSS has discovered 6 quasars at $z>5.7$
in 2870 deg$^2$ to $AB=20.2$ mag (see Fan \etal 2003). Using this as the 
normalization, and adopting $M_*=-23.9$ and $\beta_1=-2.58$ (Fan \etal 2001),
we obtain $\Phi_*=1.53\times 10^{-8}\, Mpc^{-3}\ mag^{-1}$.
The biggest uncertainty is the faint-end slope $\beta_2$,
which is essentially unknown for $z>4$. Therefore three values, $\beta_2=-1.58$,
$-2.0$ and $-2.58$, are used. The first one is the value observed at $z\simeq 2$
-- 3, the second one is the steepest value that we consider to be reasonable,
and the last one is the limiting case ($\beta_2=\beta_1$), which is considered
for illustrative purposes only. The surface densities of AGN at $z\simeq 6$ 
calculated based on these parameters are superposed in Fig. 1 as blue curves.
   
\section{Ionizing Background at $z\simeq 6$ and the Contribution of Galaxies and AGN}
 
We first need to answer the following question: in order
to keep the intergalactic hydrogen ionized at this redshift, what should be
the critical ionizing photon emission rate per unit co-moving volume
($\dot N_{cri}$)? We can then find out if the integrated production rate of
photons at $\lambda < 912$\AA\ due to either AGN or galaxies meets this
critical value.
 
   A recipe of calculating $\dot N_{cri}$ has been given in Madau, Haardt \&
Rees (1999, their Eqn. 26; hereafter MHR99):
\begin{equation}
   \dot N_{cri}(z)=10^{51.2}\ \left(\frac{C}{30}\right)\times\left(\frac{1+z}{6}\right)^3\left(\frac{\Omega_b h_{100}^2}{0.02}\right)^2\, s^{-1}\ Mpc^{-3},
\end{equation}
where $C$ is the ionized hydrogen clumping factor.
Choosing $\Omega_b h_{100}^2=0.02$, $C=30$ and $z=6$,
one finds $\dot N_{cri}=2.51\times 10^{51}\,s^{-1}\ Mpc^{-3}$. As noted by
MHR99, the time-dependent clumping factor in this expression has be written in
the form that is scaled to the value inferred at $z=5$ from the numerical
simulation of Gnedin \& Ostriker (1997), which gave $C\simeq 30$ at $z=5$.
%The actual value of $C$ might be higher, as this simulation was limited by the
%finite resolution. The uncertainty in $C$ will of course affect the value of
%$\dot N_{cri}$, and we will discuss its effect in the next section.

   The total ionizing photon production rate, $\dot N_i$, due to a given
population of objects, can be related to its differential LF, $\Phi(M)$, via
the following form (Yan 2003b) :
\begin{equation}
   \dot N_i = B\int^{M_{max}}_{-\infty} 10^{-0.4(M+25+5logD_L)}\cdot \Phi(M)\ dM\,\,\, s^{-1}\ Mpc^{-3},
\end{equation}
where $M$ is absolute magnitude (in the AB system) defined around rest-frame
1300\AA, $D_L$ is the luminosity distance (in Mpc) to the objects in question,
and $M_{max}$ is the absolute magnitude of the least luminous object that
should be used in the integration. $B$ is a constant whose value depends on the
shape of the UV-SED of these objects, which can be represented by power laws.
To be specific, $B=6.886\times 10^7\cdot D_L^2 \cdot (0.65)^\delta/(k\gamma)$,
where $\delta$ and $\gamma$ are power-law indices (in the frequency domain) for
$\lambda \leq $912\AA\, and 912\AA $\leq \lambda \leq$ 1400\AA, respectively.
The constant $k$ is the continuum discontinuity factor across the Lyman
limit at 912\AA.
 
   Following MHR99, we assume that the shape of the SED of a quasar has
power-law indices of --1.8 at $\lambda \leq$ 1050\AA\, and --0.8 at
1050\AA$\leq\lambda<$2500\AA, respectively. Using these values and the AGN LF
described in \S 2, it is found that the contribution from AGN falls far short
of the critical value $\dot N_{cri}$ for any reasonable faint-end LF slope
(see also Lehnert \& Bremer (2003) for their discussion at $z\simeq 5$). Integrated to
$M=-16.0$ mag
beyond which the existence of AGN activity becomes implausible,
their total contribution are $0.004\times 10^{51}$ and
$0.009\times 10^{51}$ s$^{-1}$ Mpc$^{-3}$ for the
faint-end slopes of $\beta_2=-1.58$ and $-2.0$, respectively. Even for the
limiting case where $\beta_2=-2.58$, the derived $\dot N_{i}$ value is only
$0.055\times 10^{51}$ s$^{-1}$ Mpc$^{-3}$.
 
   Based on 29 LBG spectra at $z\simeq 3$, Steidel et al. (2001) measured
the ratio of the continuum flux at rest-frame 1300\AA\, and at the blue side of
the Lyman limit to be $\sim$ 4.6. This is equivalent to taking
$(0.65)^\delta/k=0.217$ in our expression for $B$, and is used hereafter. Note
that by using this value, we implicitly adopt the Lyman continuum photon
escaping fraction ($f_{esc}$) of Steidel et al. (2001), which is 10--13\% in 
absolute terms. The index
$\gamma$ used in the following analysis was estimated by using a set of model
spectra of galaxies with ages of 0.1 Gyr (Bruzual \& Charlot 1993), and it
was found that this SED slope is close to an equivalent power-law index of $\gamma\simeq 1.8$.
 
    We find that the original LF estimate of Yan \etal (2002), which has a 
relatively shallow slope of $\alpha=-1.6$, cannot produce a large enough 
$\dot N_i$ that reaches $\dot N_{cri}$. This LF can account for about 67\% of 
the critical value if Eqn. 2 is integrated to $M=-16.1$ mag, but the integral
increases only very slowly towards fainter magnitudes and never meets 
$\dot N_{cri}$, even when pushed to globular-cluster-type luminosities
($M\simeq -7$ mag). Modifying our LF by adopting a luminosity evolution scheme 
of $L_*$, such as $L_*(z)=L_*(z$=3$)(1+z)^\rho/4^\rho$, will not mitigate this
problem, because the value of $\rho$ is negative in a reasonable hierarchical 
structure formation model, and thus will make $M*$ fainter and the $\dot N_i$
value even less.
 
   Thus we explore other steeper $\alpha$ values to see if $\dot N_{cri}$ can be
met at a reasonable minimum luminosity. We consider $\alpha= -1.7, -1.8, -1.9$,
and $-2.0$, where the steepest one is the critical value at which the
integral of a Schechter function diverges if the integration is carried to
infinitely low luminosity. For these slopes to meet the normalizing condition
(cumulative surface density of 1.37 per arcmin$^2$ to $m_{AB}=27.0$ mag; see
Yan \etal 2002), $\Phi_*$ has to be 4.27, 4.00, 3.74, and
$3.49\times 10^{-4} Mpc^{-3}$, respectively. We find that $\dot N_{cri}$ can be
reached at the following critical absolute magnitudes for the above slopes,
respectively: $M=$ --8.8, --14.6, --15.8, and --16.6 mag (at $z\simeq 6$ these
corresponds to apparent magnitudes of 37.9, 32.1, 30.9, and 30.1 mag,
respectively). In other words, the $least$ luminous galaxies that the Universe
should have produced by this redshift should be at these magnitude levels or
fainter in order to have a fully ionized Universe at $z\simeq 6$.
These results are shown in Fig. 2 (top panel).

\section{Discussion}

%   We have quantitatively shown that AGN can only contribute an insignificant
%fraction (less than 10\%) of the total ionizing photon budget at $z\simeq 6$.
%This leaves $normal$ galaxies as the only possible population of major
%ionizing sources at this redshift. Since the vast majority of their contribution
%comes from less luminous, but more numerous objects, $normal$ galaxies can
%indeed account for the entire ionizing photon background provided that the
%faint-end slope of their LF is sufficiently steep. As discussed in \S 3.4, this
%slope should at least be steeper than --1.6.

   The quantitative results above depend on 
three factors, namely, the clumpiness of ionized hydrogen ($C$), the
scaling factor of the LF of galaxies ($\Phi_*$), and the escaping fraction of
Lyman continuum photons ($f_{esc}$). The first one affects the value of 
$\dot N_{cri}$, while the last two affect $\dot N_{i}$. 

   The value of $\dot N_{cri}$ scales linearly with the clumping factor $C$.
In Eqn. 1, this clumping factor is scaled to its value at
$z=5$. We followed MHR99 and adopted $C=30$, which is suggested by the 
numerical simulations of Gnedin \& Ostriker (1997). Should $C$ drop to $\sim$
20, $\dot N_{cri}$ would drop by 1/3, and a faint-end slope steeper than 
$\alpha=-1.6$ would no longer be required, since now $\dot N_i$ would reach the
critical absolute magnitude at $M=-16.1$ mag. Similarly, steeper slopes, 
$\alpha=$ --1.7, --1.8, --1.9, and --2.0 would make $\dot N_{i}$ reach the
critical absolute magnitude at $M=$--17.0, --17.4, --17.7, and --17.9 mag, respectively.

   The value of $\dot N_i$ scales linearly with the scaling factor $\Phi_*$ of the LF.  
Since increasing $\dot N_i$ has the same effect as decreasing $\dot N_{cri}$,
we get exactly the same answers as above, if the scaling factor increases by a
factor of 1.5. Since $\dot N_i$ also scales linearly with $f_{esc}$ (assuming
it is not a function of luminosity), the same answers hold if $f_{esc}$ 
increases by the same factor. On the other hand, if either $\Phi_*$ or 
$f_{esc}$ drops by 1/3, a faint-end slope steeper than --1.8 is necessary to 
reionize the Universe. In this case, $\dot N_{cri}$ can be reached at $M=-12.5$
and --14.7 mag for a slope of --1.9 and --2.0, respectively. Smaller $\Phi_*$
or $f_{esc}$ requires the LF extend to fainter luminosity; however, if either
of them is smaller by a factor of two, only $\alpha=-2.0$ can meet the 
reionization requirement.

% for example, if 
%either quantity drops by a factor of two, the LF needs to extend to $M=-7.7$ 
%and $-12.7$ mag for $\alpha=-1.9$ and --2.0, respectively in order to reach
%$\dot N_{cri}$. Should either $\Phi_*$ or $f_{esc}$ be even smaller, only
%$\alpha=-2.0$ can meet the requirement.

   These values give the critical luminosity that the least luminous galaxies
should have, i.e., a lower limit to $M$. By considering the additional
constraint that the intergalactic hydrogen at slightly earlier epoch still had
a non-negligible neutral fraction, we can further improve these limits into
two-sided luminosity ranges. Cen (2003) gave a detailed modeling of the 
evolution of intergalactic
medium in the double-reionization scheme, and the neutral fraction inferred
from this modeling is about 15\% at $z\simeq 7$. The precise neutral fraction
at a specific redshift does not matter too much for our purpose; the point is
that at an earlier epoch, e.g. at $z\simeq 7$, the total ionizing photon
production rate of galaxies should not exceed the critical value such that
a significant amount of neutral hydrogen can still exist, otherwise no GP-trough
would be seen at $z\simeq 6$.

   The calculation can be performed similarly to that described in \S 3. To
first order, we can assume that there is no evolution in the LF during the 
relatively short time period from $z=6$ to 7, which is only 0.16 Gyr. At $z=7$,
$\dot N_{cri}=3.75\times 10^{51}$ s$^{-1}$ Mpc$^{-3}$ for the nominal $C$ value
of 30. Now the requirement is $N_{i}<N_{cri}$. If nothing else changes, a LF
faint-end slope of --1.8 or shallower will satisfy this criterion. Steeper 
slopes of $\alpha=$ --1.9 and --2.0 require that the LF truncates before it
reaches $M=$ --12.5 and --14.6 mag, respectively. The bottom panel of Fig. 2
demonstrates these results.

   If $C=20$, or equivalently if either $\Phi_*$ or $f_{esc}$ is increased by
a factor of 1.5, it requires that the LF truncates before it reaches $M=$ 
--12.4, --15.8, --16.9, and --17.5 mag for slopes of --1.7, --1.8, --1.9, and
--2.0, respectively (a cut-off is not required for a slope of --1.6). On the
other hand, if either $\Phi_*$ or $f_{esc}$ is decreased by 1/3, or
equivalently if $C=45$, slopes of --1.9 and --2.0 require the LF truncate
before it reaches $M=$ --4.4 and --11.8 mag. 

   Table 1 summarizes these results. We note that case a) is what we believe to
be the most plausible, given our best knowledge about the relevant parameters.
Case b) and c), on the other hand, explore a wider parameter space. However,
we point out that case b) is not very likely for several reasons.
First, the clumping factor is not likely to be as low as 20, since the nominal
value of $C=30$ obtained by Gnedin \& Ostriker (1997) might already have 
underestimated the true value because of the finite resolution of their
simulation. Second, the scaling factor of the LF, $\Phi_*$, is not likely to be
significantly higher than the value suggested in Yan \etal (2002), otherwise
all the recent searches for $z\simeq 6$ objects would have resulted
in more bright ($z$-band $m_{AB}\leq 26.0$ mag) candidates than what have been
actually observed. Third, the Lyman photon escaping fraction of star-forming
galaxies is not likely to be much higher than $\sim$ 10--13\%, given that such
a value already seems to be rather high (cf. Giallongo \etal 2002).

   The Ultra Deep Field (UDF) campaign of the ACS/HST 
(http://www.stsci.edu/hst/udf), whose data are now being taken and will be 
released to the public in February 2004, will greatly narrow down the range of 
$\alpha$. Depending on its the real value, we predict a total of 50
to 80 genuine $z\simeq 6$ objects in the UDF to its designed 10 $\sigma$ depth
of 28.4 mag.

%   We also note that a LF with a faint-end cut-off might not be desirable,
%especially if the luminosity range for the least luminous galaxies is narrow,
%since this would require the structure formation at the reionization epoch
%has a very sensitive self-regulating mechanism. With this additional
%consideration in mind, it seems that the faint-end slope of the LF is most
%likely between --1.7 and --1.8.
%We note that the Ultra Deep Field (UDF)
%campaign of the ACS/HST, whose data are now being taken and will be released to
%the public in early 2004, will narrow down the range of $\alpha$. Depends on
%its real value, our LF predicts a total number of 50 to 80 genuine $z\simeq 6$
%objects in the UDF to its designed depth of $m_{AB}(F850LP)=28.4$ mag. 

\section{Summary}

   In this paper we investigate the major sources of reionization at
$z\simeq 6$ based on the best available observational constraints.
Using the LF estimate proposed for galaxies at this redshift
(Yan \etal 2002) as the starting point, we find that $normal$ galaxies can
account for the entire reionizing background, provided that the faint-end slope
of their LF is steep enough and that the LF extends to sufficiently low 
luminosity. We explore a range of faint-end slopes, from $\alpha=-1.6$, a value
that the LF has at $z\simeq 3$, to $\alpha=-2.0$, the critical value at
which the integration of a Schechter LF diverges. We find that dwarf galaxies,
rather than their more luminous counterparts or AGN, produced the vast
majority of the reionizing photons at $z\simeq 6$. Using the best estimate of
the relevant parameters, we find that the faint-slope of the LF of galaxies at
$z\simeq 6$ should be steeper than --1.6, if the adopted $\Phi_*$ holds, or
as steep as $\alpha=-2.0$ if $\Phi_*$ is lower by a factor of two.
In addition, we show that the least luminous galaxies should be
fainter than a certain critical luminosity, whose exact value depends on the 
actual faint-end slope. Furthermore, by requiring that there still was a 
non-negligible neutral hydrogen fraction at slightly earlier epochs
($z\simeq 7$), we point out that the LF should truncate at a certain luminosity
threshold, whose value again depends on the actual faint-end slope.
The HST UDF data will be able to narrow down the range of this important 
parameter, whose precise value is obviously important for planning the
performance of the James Webb Space Telescope.

\acknowledgments
Funding from NASA/JWST Grant NAG5-12460 is acknowledged. The authors would like
to thank Dr. Mark Dickinson for pointing out the hidden step in the ACS pipeline
reduction that led to the improper zeropoint used in Yan \etal (2003a). We
also thank Drs. Piero Madau and Joe Silk for helpful discussions. We thank
the referee for the helpful comments.

\clearpage
\begin{figure}
\epsscale{.90}
\plotone{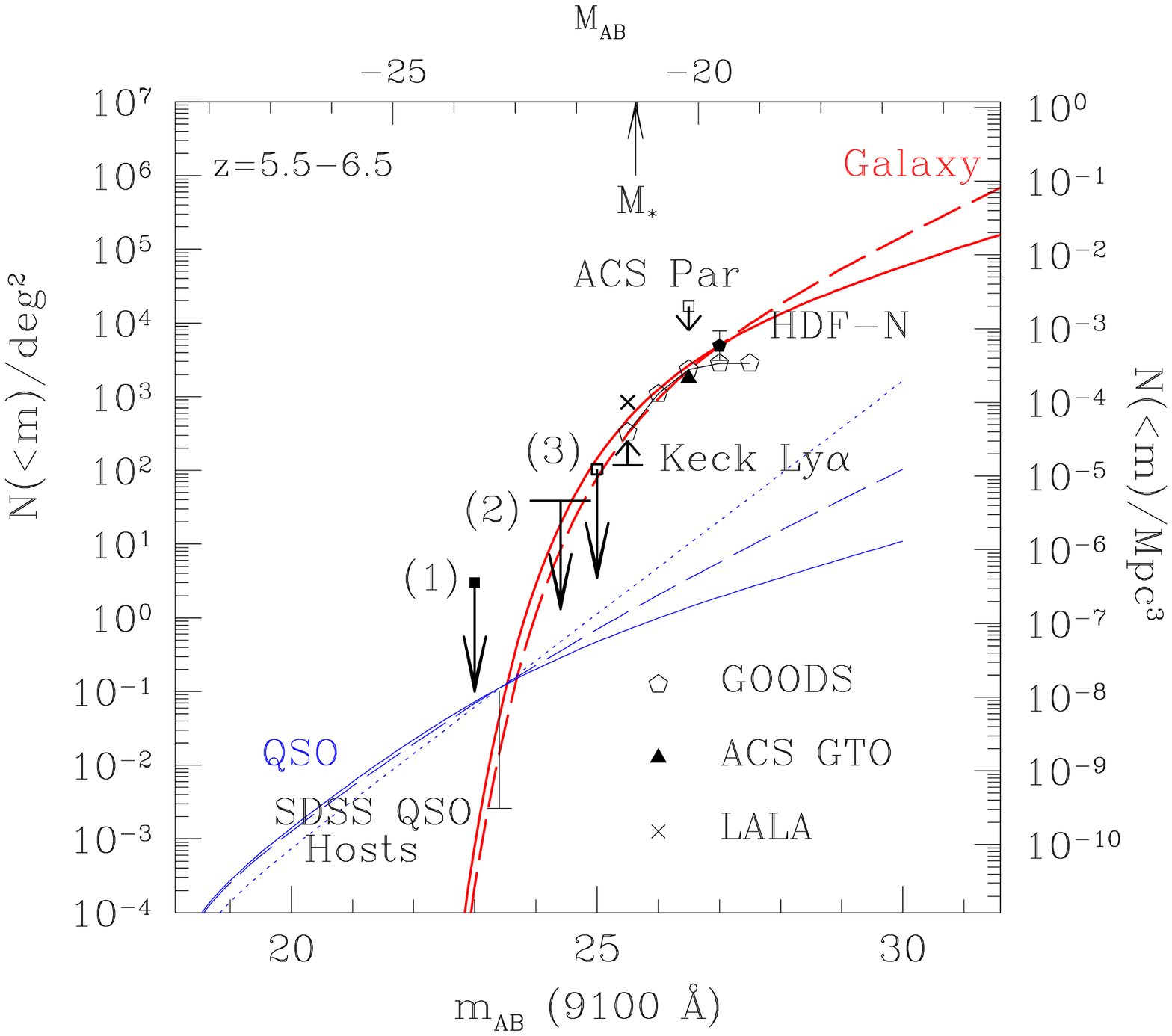}
\caption{The best estimate of the LF of galaxies at $z\simeq 6$ (red curves),
based on Yan \etal (2002), agrees with all known observational constraints.
Two faint-end slopes, $\alpha=-1.6$ (solid line) and $-2.0$ (long dashed line),
are shown. Only new constraints are discussed here. The upper limits marked
(1) and (2) are derived from a degree-sized intermediate-band survey
(Yan \etal 2003, in preparation), while the one marked (3) is derived based on
the deep Subaru data of Capak \etal (2003). The cross is derived from the LALA
survey (Rhoads \etal 2003). The solid triangle is based on Bouwens \etal (2003).
The open pentagons connected by line are based on the results of the GOODS 
(Dickinson \etal 2003). The upper limit based on the ACS parallel observation
of Yan \etal (2003a) has been corrected to $AB=26.5$ mag to account for the
proper magnitude zeropoint. This LF predicts that 50--80 $z\simeq 6$ objects 
will be found by the UDF campaign. For comparison, the blue curves
show the expected number density of quasars at $z\simeq 6$, derived by using a
double-power law LF and the normalization based on the SDSS $z\simeq 6$ quasar
sample. Three different faint-end slopes are assumed, $\beta_2=-1.58$ (solid
curve), --2.0 (dashed curve), and --2.58 (dotted line). A truncation at $M=-16$
mag is applied, as beyond this point the existence of significant AGN activity
is implausible.
}
\end{figure}

\begin{figure}
\epsscale{.90}
\plotone{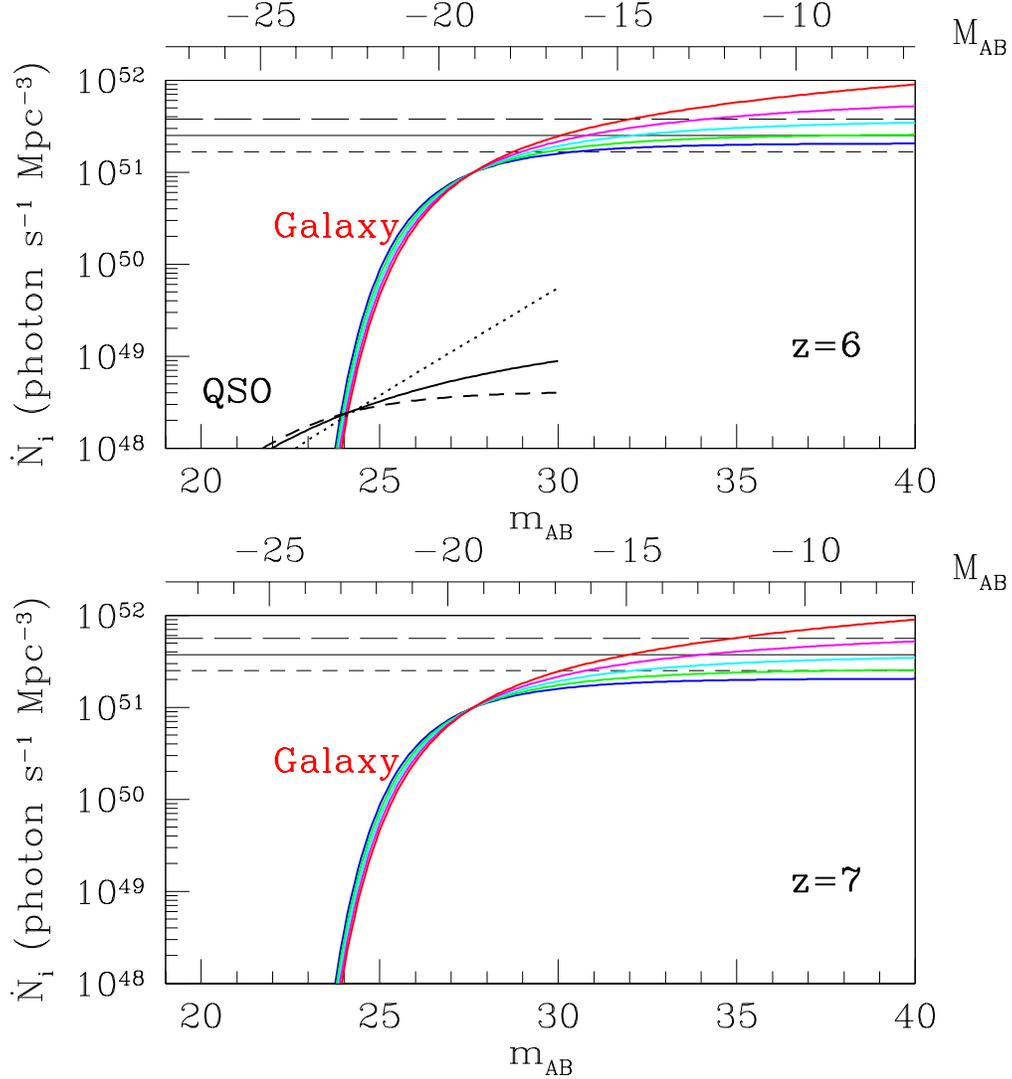}
\caption{(Top) At $z\simeq 6$, the reionizing photon production rate due to 
galaxies should meet the critical value in order to keep the Universe fully
ionized. The solid, long-dashed and short-dashed horizontal lines are the
critical values if the clumping factor (at $z=5$) $C=30$, 45 and 20,
respectively.  The contribution from galaxies is shown in color, assuming 
different faint-end slopes of their LF: --1.6 (blue), --1.7 (green), --1.8
(cyan), --1.9 (magenta), and --2.0 (red). Galaxies can account for the entire
reionizing photon budget provided that their LF is sufficiently steep
($\alpha\leq -1.7$), and extends to sufficiently low luminosity. 
%i.e., there is a minimum luminosity that the least luminous galaxies should at least reach. 
Quasars can only contribute
an insignificant fraction of the ionizing photons, as the black curves show:
the solid, long-dashed and dotted lines are for faint-end slopes of --1.58, 
--2.0 and --2.58, respectively. (Bottom) The reionizing photon production rate
due to galaxies at a slightly earlier epoch should be smaller than the
corresponding critical value, as there was still a significant fraction of
neutral hydrogen. The case is shown for $z\simeq 7$. Legends are the same as
in the top panel. The LF of galaxies should cut-off at a certain
luminosity, i.e., there is a luminosity threshold for the $least$ luminous
galaxies at this redshift, below which the LF cannot continue or it would leave
no neutral hydrogen at $z\simeq 7$ and no GP-trough in SDSS quasars at $z > 6$.
}
\end{figure}

\begin{table}
\caption{The luminosity range of the least luminous galaxies that should exist at reionization}
\begin{center}
\begin{tabular}{ccccc}\tableline \tableline

$\alpha$ & $\Phi_*^\dagger$ & $M_{min}$\tablenotemark{a} & $M_{min}$\tablenotemark{b} & $M_{min}$\tablenotemark{c} \\ \tableline
--1.6 & 4.55 & N/A & $-16.1\leq M_{min}$  & N/A \\
--1.7 & 4.27 & $-8.8\leq M_{min}$ & (--17.0, --9.1) & N/A \\
--1.8 & 4.00 & $-14.6\leq M_{min}$ & (--17.4, --14.6) & N/A \\
--1.9 & 3.74 & (--15.8, --12.6) & (--17.7, --15.9) & (--12.5, --4.4) \\
--2.0 & 3.49 & (--16.6, --14.7) & (--17.9, --16.6) & (--14.7, --11.8) \\
\tableline

\end{tabular}
\end{center}
\tablenotetext{\dagger} {Nominal scaling factor of the LF (in unit of $10^{-4} Mpc^{-3}$) by forcing an accumulative number density of 1.37 per arcmin$^2$ to
$m_{AB}=27.0$ mag (Yan \etal 2002).}
\tablenotetext{a} {For the case where $C=30$, and $\Phi_*$ and $f_{esc}$ have
their nominal values (see text). The minimum absolute magnitudes are referred
to rest-frame wavelength at around 1300\AA, and are in the AB system.}
\tablenotetext{b} {For $C=20$, or equivalently either $\Phi_*$ or $f_{esc}$ 
increased by a factor of 1.5.}
\tablenotetext{c} {For $C=45$, or equivalently either $\Phi_*$ or $f_{esc}$
decreased by 1/3. If $\Phi_*$ decreases by a factor of two, only $\alpha=-2.0$
can meet the reionization requirement.}

\end{table}

\end{document}